\documentclass[%
 reprint,
superscriptaddress,
 amsmath,amssymb,
prb,
]{revtex4-2}

\usepackage{graphicx}
\usepackage{dcolumn}
\usepackage{bm}
\usepackage{hyperref}
\usepackage{ulem}

\hypersetup{
	colorlinks=true,
	linkcolor=blue,
	citecolor=blue,
	filecolor=blue,      
	urlcolor=blue,
}
\usepackage{hypcap}
\newcommand{\commented}[1]{}

\usepackage[dvipsnames]{xcolor} 
\begin{document}

\preprint{APS/123-QED}


%
%
%
\title{Theory of Momentum-Resolved Electron Energy-Loss Spectra of Coupled Phonon and Magnon Excitations}

%
%

%
%
\author{Jos\'e \'Angel Castellanos-Reyes}
\email{angel.castellanos.research@gmail.com}
\affiliation{Department of Physics and Astronomy, Uppsala University, Box 516, 75120 Uppsala, Sweden}

\author{Ivan P. Miranda}
\affiliation{Department of Physics and Electrical Engineering, Linnaeus University, SE-39231 Kalmar, Sweden}
\author{Paul M. Zeiger}
\affiliation{Department of Physics and Astronomy, Uppsala University, Box 516, 75120 Uppsala, Sweden}
\affiliation{Materials Sciences and Engineering Department, University of Washington, Seattle, WA 98115, USA}

\author{Anders Bergman}
\affiliation{Department of Physics and Astronomy, Uppsala University, Box 516, 75120 Uppsala, Sweden}

\author{J\'an Rusz}
\email{jan.rusz@physics.uu.se}
\affiliation{Department of Physics and Astronomy, Uppsala University, Box 516, 75120 Uppsala, Sweden}

\date{\today}
%
%
\begin{abstract}

%
%
We develop a theory of momentum-resolved electron energy-loss spectra in the scanning transmission microscope (STEM-EELS) that captures the effects of coupled phonon and magnon excitations within a unified formalism, and apply it to body-centered cubic iron at 300~K. By advancing the Time Autocorrelation of Auxiliary Wavefunctions (TACAW) method to incorporate atomistic spin-lattice dynamics (ASLD), we simulate the EELS signal, including phonon-magnon interaction effects, dynamical diffraction, and multiple scattering. 
Our results reveal non-additive spectral features arising from phonon–magnon coupling, hybridization, and energy shifts, and further allow estimation of the electron dose required to detect magnon scattering under optimized detector conditions. 
\end{abstract}
\maketitle

%
%

Resolving phonon and magnon excitations at the nanoscale is essential for understanding thermal, electric, magnetic, optical, mechanical, and quantum phenomena in real materials~\cite{Barman2021,Flebus_2024,cahill,gaoreview2025}. Since the first detection of phonons using electron energy-loss spectroscopy (EELS) in a scanning transmission electron microscope (STEM) in 2014~\cite{krivanek2014nature}, STEM-EELS has emerged as a premier technique for vibrational spectroscopy with (sub-)nanometer spatial resolution \cite{gaoreview2025,haas2024,lagos2022advances}. Only recently, however, have magnons---the quanta of spin waves~\cite{Eriksson2017}---been directly observed at the nanoscale in the bulk using STEM-EELS~\cite{naturemagnoneels}, made possible by advances in monochromation \cite{krivanek2014nature,krivanek2019ultramicroscopy}, detector design \cite{plotkin-swing_hybrid_2020}, and theoretical modeling~\cite{tacawpaper}.

Despite these experimental breakthroughs, the theoretical STEM-EELS models continue to treat phonons and magnons as independent scattering channels~\cite{zeiger2023lessons,mendis2021,mendis2022,Lyon2021PRB,castellanos_unveiling_2023,tacawpaper,naturemagnoneels,julio2024,gaoreview2025,haas2024}. This simplification limits their applicability to real materials, where both excitations coexist and interact. This limitation is particularly severe for interpreting magnon STEM-EEL spectra, where magnetic signals are often buried under a phonon background three or more orders of magnitude stronger~\cite{Lyon2021PRB,mendis2022,castellanos_unveiling_2023,tacawpaper,naturemagnoneels}. Even in energy regions where phonon and magnon branches are well separated, the phonon background may still obscure the magnon signal~\cite{naturemagnoneels}, complicating detection and interpretation~\footnote{See Supplementary Note 4 of Ref.~\cite{naturemagnoneels}}. Furthermore, phonon-magnon coupling can modify the temperature dependence of phonon peak shifts detectable in STEM-EELS measurements, and a theoretical description of this effect naturally requires incorporating coupling between lattice and spin degrees of freedom \cite{reifsnyderDetectingMagnonphononCoupling2026}.

The Time Autocorrelation of Auxiliary Wavefunctions (TACAW) method~\cite{tacawpaper} has recently established a robust framework for simulating spatially- and momentum-resolved ultra-low-energy-loss\footnote{In STEM-EELS, \textit{ultra-low-energy-loss} refers to excitations with energies in the millielectronvolt range.} STEM-EELS signals by propagating auxiliary electron-beam wavefunctions through a sequence of atomic configurations sampled along a finite-temperature trajectory. Until now, TACAW has been applied separately to phonons, using molecular dynamics (MD)~\cite{tacawpaper,zeigermandm,pfeifer2025,hoglund2025,gaoreview2025}, and to magnons, using atomistic spin dynamics (ASD)~\cite{tacawpaper,naturemagnoneels}.

In this work, we reconceptualize TACAW by integrating it with first-principles-driven atomistic spin–lattice dynamics (ASLD)~\cite{Eriksson2017,asldpaper}, creating a method that captures vibrational and magnetic excitations on equal footing---something previously inaccessible within TACAW implementations.
This ASLD–TACAW framework naturally incorporates the real-space interplay between atomic motion and spin dynamics and reveals the resulting momentum-space redistribution of spectral weight---key signatures that standard TACAW formulations could not resolve.

In this context, body-centered cubic iron (bcc Fe) at room temperature provides an ideal test case. It is a prototypical ferromagnet for which magnon STEM and STEM-EELS simulations have been reported~\cite{Lyon2021PRB,mendis2022,castellanos_unveiling_2023,tacawpaper}, with well-characterized phonon and magnon dispersions~\cite{Peterkohn}, known spin-phonon coupling~\cite{Miranda2025}, and a high Curie temperature. 
Moreover, recent \textit{ab initio} studies~\cite{asldpaper,Miranda2025} identify Fe as the elemental 3\textit{d} ferromagnet exhibiting the strongest sensitivity of the spin–lattice couplings to perturbations from the collinear magnetic state. Its nearest-neighbor interactions are dominated by short-ranged, mixed $e_g$–$t_{2g}$ orbital contributions, making it a simple yet sensitive platform to investigate the influence of coupled phonon–magnon dynamics on the STEM-EELS signal.


%
%

Therefore, in this Letter, we develop a STEM-EELS theory that treats phonon and magnon excitations on equal footing---within a single finite-temperature trajectory---by incorporating ASLD into TACAW. As a proof of concept, we apply the method to bcc Fe at 300~K, employing the first-principles parametrization of Refs.~\cite{Pankratova2024,Miranda2025}. 

The theory employs the TACAW method, which computes $I(\mathbf{q}, E)$---the STEM-EELS scattering cross-section as a function of momentum transfer $\mathbf{q}$ and energy $E$---as the time ($t$) to energy Fourier transform of the autocorrelation of the auxiliary wavefunction $\psi(\mathbf{q},t)$ obtained from multislice 
propagation through the time-evolving atomic and magnetic configurations generated along the ASLD trajectory~\cite{tacawpaper}. 
Incorporating ASLD provides these time-resolved configurations as a finite-temperature trajectory of atomic positions and magnetic moments, capturing the coupled spin–lattice dynamics of the system.
We use the Pauli multislice method~\cite{edstrom_elastic_2016,edstrom_magnetic_2016,Lyon2021PRB}---a magnetic-field-aware multislice formulation---to propagate the electron beam's wavefunction through each snapshot. 
The overall approach remains computationally efficient and scalable, and provides predictive access to momentum-resolved STEM-EELS signals at finite temperature under realistic experimental conditions.



%
%

\begin{figure*}[t!]
    \centering
    \includegraphics[width=\linewidth]{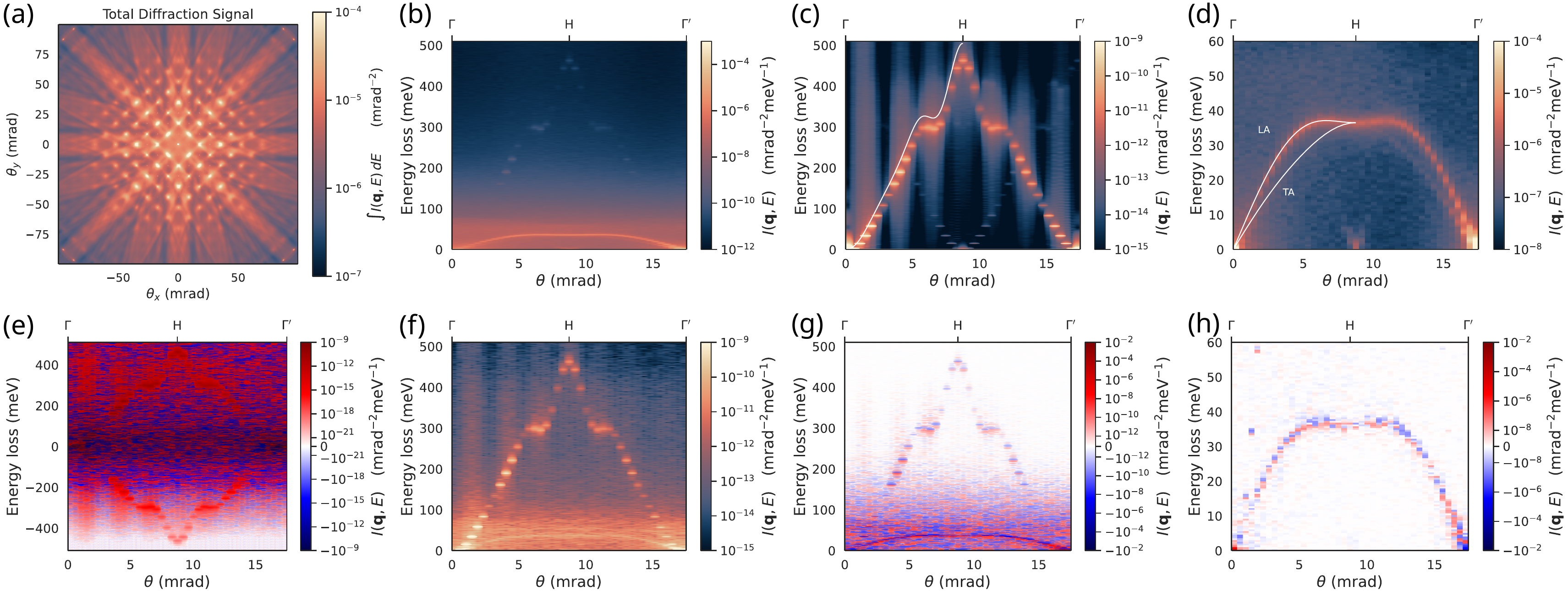}
    \caption{\textbf{Momentum-resolved EELS simulations from ASLD–TACAW for bcc Fe at 300\,K in the [001] zone axis.}
    \textbf{(a)} Total (elastic+inelastic) diffraction intensity, $\int I(\mathbf{q},E)\,dE$, on a logarithmic scale, showing Bragg peaks and diffuse scattering.
    \textbf{(b)} CPLED: fully coupled ASLD–TACAW momentum-resolved EELS, $I(\mathbf{q},E)$, including phonons, magnons, and their mutual coupling.
    \textbf{(c)} MG-ASLD: magnon-only $I(\mathbf{q},E)$ obtained by freezing atomic positions in the ASLD trajectory, showing a broad magnon band extending up to $\sim 500$\,meV (adiabatic magnon dispersion overlaid in white).
    \textbf{(d)} PH-ASLD: phonon-only $I(\mathbf{q},E)$ obtained by freezing magnetic moments, revealing LA and TA branches below $50$\,meV (adiabatic dispersions overlaid in white).
    \textbf{(e)} DIFF-PH = CPLED $-$ PH-ASLD (signed residual), plotted for both energy-loss and gain regions (negative energies) on a symmetric diverging scale, highlighting coupling-induced redistributions relative to the phonon background.
    \textbf{(f)} $|\text{DIFF-PH}|$, approximating a background-subtracted experiment where a phonon baseline is removed to isolate the magnetic component and coupling effects.
    \textbf{(g)} DIFF-UNC = CPLED $-$ UNCOUPLED, where UNCOUPLED is the additive sum of MD-TACAW phonon and ASD-TACAW magnon spectra, showing sign-alternating residuals that track the magnon dispersions from $\sim 150$\,meV to $\sim 500$\,meV.
    \textbf{(h)} DIFF-UNC in the phonon energy range, exhibiting sign-alternating lobes localized around the LA branch.
}
    \label{fig:asld-eels}
\end{figure*}

Figure~\ref{fig:asld-eels} summarizes the momentum-resolved EELS signal computed using ASLD–TACAW for bcc Fe at 300~K for a 200~kV parallel beam illumination in the $[001]$ zone axis~\footnote{Following the methodology outlined in Refs.~\cite{castellanos_unveiling_2023,tacawpaper}, we considered $20 \times 20 \times 70$ supercell of bcc iron with a lattice parameter of 2.8665~\AA{} (i.e., with a thickness of 20~nm) and magnetic moments of 2.23~$\mu_B$/atom (where $\mu_B$ is the Bohr magneton), computed \textit{ab initio} along with exchange interactions using the scalar-relativistic \textsc{RS-LMTO-ASA} method~\cite{Frota-Pessoa1992}. Using the \textsc{UppASD} package \cite{UppASD}, we performed ASLD dynamics by solving coupled Langevin equations of motion for the spin-lattice system \cite{asldpaper} with a time step of 0.1~fs, a Gilbert damping parameter $\alpha=10^{-4}$, and a lattice damping of $\nu\sim 2$~ps$^{-1}$, which corresponds to $\gamma= 2\times10^{-13}$ kg/s in the notation of Ref.~\cite{Ma}. The \textit{ab initio} force constants and spin-lattice coupling (SLC) parameters were taken from Ref. \cite{Pankratova2024}. (see Ref. \cite{Miranda2025} for a more detailed calculation and analysis of the SLC parameters.) The thermalization phase consisted of $10^5$ steps, and was followed by a 0.2~ns trajectory at 300~K for generating snapshots. In total, 50{,}000 snapshots were generated with a $\Delta t=4$~fs, enabling exploration of magnon frequencies up to 125~THz ($\sim 517$~meV). The exit wave functions were computed using the Pauli multislice method on a numerical grid of $1000 \times 1000$ with $140$ slices across the supercell's thickness. A 200-kV parallel-beam-illumination electron probe propagating along the [001] direction was employed. We have used the parametrized magnetic vector potential developed in Ref.~\cite{lyon_parameterization_2021}. To obtain $I(\mathbf{q},E)$, we averaged over 241 sets of 2000 consecutive snapshots (i.e., $T=8.0$~ps), mutually offset by 200 snapshots. We considered a constant magnetic field of 1 T applied in the $[001]$ direction to account for the objective lens of the microscope. All momentum transfers $\mathbf{q}$ lie in the $xy$ plane, so the horizontal path used in Fig.~\ref{fig:asld-eels} follows $\Gamma\!-\!H\!-\!\Gamma$ in the $(q_x,q_y)$ plane.}. For clarity, we introduce the following notation for the different simulations used in this work: 
\textbf{CPLED} denotes the fully coupled ASLD–TACAW calculation, where both atomic positions and magnetic moments evolve dynamically in time and spin–lattice coupling is active. 
\textbf{PH-ASLD} refers to an ASLD trajectory in which the magnetic moments are \textit{a posteriori} frozen (magnons suppressed), yielding a phonon-only EELS signal within the same spin–lattice Hamiltonian. 
\textbf{MG-ASLD} denotes the complementary case in which the atomic positions are frozen while the spins evolve, isolating the magnon contribution. 
As a reference to earlier work, we define \textbf{UNCOUPLED} as the sum of phonon-only spectra obtained from standalone MD-TACAW and magnon-only spectra obtained from standalone ASD-TACAW, without mutual coupling between spins and lattice. 
From these, we construct two residual spectra: 
\textbf{DIFF-PH}~$\equiv$~CPLED~$-$~PH-ASLD, highlighting magnetic and coupling contributions relative to an ASLD phonon background, and 
\textbf{DIFF-UNC}~$\equiv$~CPLED~$-$~UNCOUPLED, which exposes non-additive spectral redistributions that cannot be captured by an additive combination of independent phonon and magnon channels.

Figure~\ref{fig:asld-eels}(a) shows the total electron diffraction intensity, $\int I(\mathbf{q}, E)\,dE$, in logarithmic scale, where the Bragg peaks and diffuse inelastic background are visible. 
The CPLED ASLD–TACAW $I(\mathbf{q}, E)$ along the horizontal $\Gamma\!-\!H\!-\!\Gamma$ cut, shown in Fig.~\ref{fig:asld-eels}(b), reveals strong phonon scattering below 50~meV and weaker dispersive features at higher energies that are consistent with finite-temperature spin-wave excitations exhibiting known Kohn anomalies \cite{Peterkohn,Halilov_1997}.

To disentangle the different energy-loss channels, we compare CPLED with the partial ASLD spectra MG-ASLD and PH-ASLD. 
Figure~\ref{fig:asld-eels}(c) shows MG-ASLD, obtained by freezing the atomic positions and allowing only the magnetic moments to evolve, which yields a broad magnon band extending up to $\sim$500~meV (adiabatic magnon dispersion in white). 
Figure~\ref{fig:asld-eels}(d) shows PH-ASLD, obtained by freezing the magnetic moments and allowing only the lattice to evolve, where the acoustic branches (TA, LA; adiabatic dispersions overlaid in white) are clearly resolved below 50~meV.

Figures~\ref{fig:asld-eels}(e)–\ref{fig:asld-eels}(h) then isolate the effect of spin–lattice coupling. 
First, we construct DIFF-PH = CPLED $-$ PH-ASLD. 
The signed DIFF-PH residual (Fig.~\ref{fig:asld-eels}(e), plotted on a symmetric diverging logarithmic scale) emphasizes the momentum-dependent magnetic and coupling-induced redistribution on top of the phonon background, while the unsigned $|\text{DIFF-PH}|$ magnitude (Fig.~\ref{fig:asld-eels}(f)) approximates a background-subtracted experiment in which a phonon baseline is removed to reveal the magnetic component and coupling-induced reweighting. 
This is the type of signal reported in Ref.~\cite{naturemagnoneels} for NiO, via an approximate fit-based background removal.
Next, Figs.~\ref{fig:asld-eels}~(g) and \ref{fig:asld-eels}~(h) show DIFF-UNC = CPLED $-$ UNCOUPLED, where UNCOUPLED is the additive baseline obtained by summing phonon-only MD-TACAW and magnon-only ASD-TACAW spectra without mutual coupling. 
The map in Fig.~\ref{fig:asld-eels}~(g) shows sign-alternating, structured residuals that follow the magnon dispersions from $\sim$150~meV to $\sim$500~meV, demonstrating that coupling induces a non-additive redistribution of spectral weight along the magnon band. 
The phonon-window zoom in Fig.~\ref{fig:asld-eels}~(h) reveals sign-alternating lobes localized predominantly around the longitudinal acoustic (LA) branch. 
This is a hybridization fingerprint expected from exchange-striction–mediated magnon–phonon mixing in Fe: intensity is transferred onto mixed modes where energy and momentum match, with adjacent depletion on either side of the branch~\cite{asldpaper,Miranda2025}. 
These qualitative markers---branch selectivity, momentum dependence, and sign-alternating lobes---are consistent with atomistic spin–lattice theory for Fe and with the strongly $\mathbf{q}$-dependent interactions reported in \textit{ab initio} studies~\cite{asldpaper,Miranda2025}.

\begin{figure*}[t!]
    \centering
    \includegraphics[width=\linewidth]{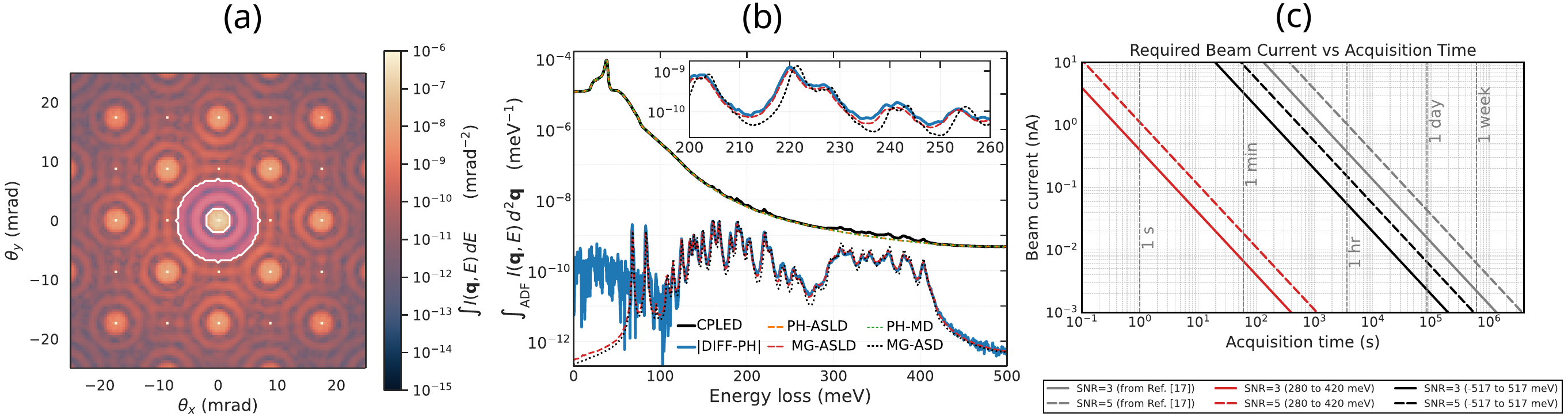}
\caption{
    \textbf{Magnetic signal detectability from detector-integrated EELS.}
    \textbf{(a)} Energy-integrated angular distribution of the MG-ASLD (magnon-only ASLD) signal in the diffraction plane, highlighting an ADF detector with inner and outer collection angles of 2~mrad and 7~mrad, following the optimal detection geometry reported in Ref.~\cite{castellanos_unveiling_2023}.
    \textbf{(b)} Momentum-resolved EELS spectra, $I(\mathbf{q},E)$, integrated over the ADF detector, showing CPLED, PH-ASLD, MG-ASLD, and $|\text{DIFF-PH}|$, together with the uncoupled reference spectra PH-MD (phonons from MD-TACAW) and MG-ASD (magnons from ASD-TACAW) that define UNCOUPLED. The inset magnifies the 200--280~meV magnetic window, where small but systematic meV-scale energy shifts between the coupled (MG-ASLD) and uncoupled (MG-ASD) spectra become visible.
    \textbf{(c)} Required beam current versus acquisition time to achieve SNR = 3 and SNR = 5 for the magnon signal integrated either over the magnon energy window (280--420~meV, red curves) or over the full simulated energy range (black curves). Gray lines indicate the corresponding estimates from Ref.~\cite{castellanos_unveiling_2023}. 
}
    \label{fig:snr-feasibility}
\end{figure*}

Figure~\ref{fig:snr-feasibility} addresses the detectability of the magnetic signal. We employ an annular dark-field (ADF) detector with inner and outer collection angles of 2 and 7~mrad, respectively—identified in Ref.~\cite{castellanos_unveiling_2023} as optimal for capturing magnon diffuse scattering in bcc Fe.
Figure~\ref{fig:snr-feasibility}(a) shows the angular distribution of the MG-ASLD intensity in the diffraction plane, guiding the choice of detector geometry for maximizing magnetic sensitivity. The selected ADF region is marked by the white boundary enclosing the blue–to–magenta color scale.

Figure~\ref{fig:snr-feasibility}(b) displays the resulting momentum-resolved EELS signal, $I(\mathbf{q},E)$, integrated over this ADF detector. We compare CPLED, PH-ASLD, MG-ASLD, and $|\text{DIFF-PH}|$ with the uncoupled reference spectra used to define UNCOUPLED, namely PH-MD (phonons from standalone MD-TACAW) and MG-ASD (magnons from standalone ASD-TACAW). As expected, PH-ASLD dominates the low-energy region, whereas both $|\text{DIFF-PH}|$ and MG-ASLD exhibit detectable magnetic contributions above $\sim$150~meV.

To further disentangle the influence of phonon–magnon coupling, we compare the ASLD-based spectra with their uncoupled MD- and ASD-TACAW counterparts. Below 150~meV, CPLED, PH-ASLD, and PH-MD nearly coincide, confirming that the signal is predomidominantly phononic.
Above 150~meV, magnetic signals emerge: MG-ASLD and MG-ASD agree qualitatively, though up to about 300~meV the former shows slightly higher intensity, suggesting that coupling somewhat strengthens the magnetic scattering. 
At higher energies ($>$300~meV), MG-ASD displays sharper minima and maxima, indicating that coupling introduces a weak damping to the spin dynamics. 
A closer view of the magnetic signals, shown in the inset of Fig.~\ref{fig:snr-feasibility}(b), reveals small but systematic meV-scale shifts between the MG-ASLD and MG-ASD peak positions. These shifts constitute direct evidence of the magnon–phonon renormalization expected from spin–lattice coupling.
Notably, CPLED lies above the additive UNCOUPLED baseline (see the blue curve below 150~meV), showing non-additive spectral redistribution arising from the interplay of phonons and magnons.

Finally, Fig.~\ref{fig:snr-feasibility}(c) quantifies the STEM beam current required to reach signal-to-noise ratios (SNR) of 3 or 5 as a function of acquisition time for the detector-integrated magnetic signal. The red curves correspond to integration over the magnon energy window (280–420~meV), while the black curves represent integration over the full simulated range ($-517$ to $+517$~meV), i.e., without energy filtering. For reference, the estimates from Ref.~\cite{castellanos_unveiling_2023} are included as gray lines.

Under the conditions considered here (200~kV beam energy, $\sim$20~nm sample thickness, and ADF collection from 2 to 7~mrad), magnon EELS signals in bcc Fe at 300~K are experimentally accessible under demanding but realistic conditions. Restricting the analysis to the magnon window improves the SNR in the optimal ADF detector by roughly three orders of magnitude for a fixed beam current, reducing the required acquisition time from days to hours—or even minutes---depending on the targeted SNR. Note, though, that bcc Fe offers a particularly convenient large separation of the phonon and magnon modes, which is often not the case in other materials. Depending on specific experimental parameters---such as sample thickness, beam energy, detector sensitivity, and noise levels---acquisition times may range from minutes to several hours or days.

%
%

Our results demonstrate that the ASLD-TACAW framework enables momentum-resolved simulations of EELS spectra in systems with coupled phonon and magnon excitations. By treating both vibrational and magnetic modes within a unified dynamical formalism---including their mutual interactions and interference---we overcome the inherent limitations of additive models by capturing spectral interference and redistribution, thereby enabling quantitative isolation of weak magnetic signals within dominant phonon backgrounds. The inclusion of dynamical diffraction, thermal disorder, and inelastic channel coupling provides realistic signal shapes and intensities that can guide experimental acquisition strategies.

Applied to bcc Fe, the method successfully reproduces distinct phonon and magnon features and reveals non-additive residuals attributable to coupling. By combining signal subtraction with angular and spectral integration, we identify experimental conditions under which magnetic excitations become detectable, even when buried under several orders of magnitude more intense phonon background. These insights are critical for designing future STEM-EELS experiments targeting spin excitations in complex systems. Beyond bcc Fe, this approach can be readily extended to other magnetic materials, paving the way for predictive simulations of spin–phonon interplay in low-dimensional magnets~\cite{gupta2022fundamentals}, multiferroics~\cite{Spaldin2019}, and materials exhibiting ultrafast magnetization dynamics~\cite{Bigot2013} where spin-lattice interactions play a central role.


%
%

Within the present implementation, the lattice dynamics are treated within the harmonic approximation and higher-order magnon–phonon interactions are neglected, which may become important at elevated temperatures or for large-amplitude excitations. 
Future extensions of the ASLD–TACAW method may include anharmonic lattice dynamics~\cite{Simak2011}, higher-order spin interactions~\cite{asldpaper,Eriksson2017}, and machine-learning-based spin–lattice potentials~\cite{Rinaldi2024}, as well as time-resolved simulations to capture ultrafast magnetization dynamics~\cite{caruso20252roadmapultrafastdynamics}. 
Such developments would broaden the applicability of the framework to complex magnetic materials with strong spin–lattice coupling and provide quantitative guidance for next-generation time-resolved STEM-EELS experiments.


%
%
\begin{acknowledgments}
We acknowledge the support of the Swedish Research Council (grant nos.\ 2021-03848 and 2024-06617), Olle Engkvist's foundation (grant no.\ 214-0331), STINT (grant no. CH2019-8211), Knut and Alice Wallenbergs' foundation (grant no.\ 2022.0079), the Carl Trygger foundation, eSSENCE and Uppsala University’s Ai4Research center. The simulations were enabled by resources provided by the National Academic Infrastructure for Supercomputing in Sweden (NAISS) at NSC Centre partially funded by the Swedish Research Council through grant agreement no.\ 2022-06725. We are grateful to Peter M. Oppeneer and Rafael M. Vieira for stimulating discussions.
\end{acknowledgments}

\section*{Author Contributions}

J.\ Á.\ C.-R. and J.\ R.: conceptualization and research design.  
I.\ P.\ M. and A.\ B.: computation of ASLD input parameters, including determination of spin–lattice couplings (methodology, data curation).  
J.\ Á.\ C.-R.: execution of all ASLD and ASLD–TACAW simulations and data analysis.  
All authors: discussion and interpretation of results; manuscript writing, review and editing.


%
%


%
\bibliographystyle{apsrev4-2}

\bibliography{references}


\end{document}